\documentclass[aps,prd,reprint,nofootinbib, superscriptaddress]{revtex4-1}
\usepackage{amsmath,amssymb, amsthm,amstext}
\usepackage{natbib}
\usepackage{graphicx}
\usepackage{color}
\usepackage{array, enumerate}
\usepackage{bm}
\usepackage{multirow}
\usepackage[breaklinks,colorlinks,citecolor=blue]{hyperref}

\def\nn{\nonumber}

\def\ct{\cos\theta}
\def\sst{\sin^2\theta}

\def\Ap{A_\phi}
\def\Ar{A_{\phi,r}}
\def\Arr{A_{\phi,rr}}
\def\Am{A_{\phi,\mu}}
\def\Amm{A_{\phi,\mu\mu}}

\def\be{\begin{equation}}
\def\ee{\end{equation}}
\def\ben{\begin{eqnarray}}
\def\een{\end{eqnarray}}
\def\WH{\Omega_{\rm H}}

\def\AHE{A_\phi^{\rm HE}}
\def\AEE{A_\phi^{\rm EE}}

\begin{document}
\title{Magnetosphere structure of a Kerr black hole:
marginally force-free equatorial boundary condition}
\author{Zhen Pan}
\email{zpan@perimeterinstitute.ca}
\affiliation{Perimeter Institute for Theoretical Physics, Waterloo, Ontario N2L 2Y5, Canada}

\date{\today}

\begin{abstract}
    The role of equatorial boundary condition in the structure of
    a force-free black hole magnetosphere was rarely discussed, since previous studies have been focused on the
    field lines entering the horizon. However,
    recent high-accuracy force-free electrodynamics (FFE) simulations \cite{East2018}
    show that there are both field lines entering the horizon and field lines ending up on
    the equatorial current sheet within the ergosphere for asymptotic uniform field configuration.
    For the latter field lines, the equatorial boundary
    condition is well approximated being marginally force-free, i.e., $B^2-E^2\approx 0$, where $B$ and
    $E$ are the magnetic and electric field strength, respectively. In this paper, we revisit the uniform field
    solution to the Kerr BH magnetosphere structure and investigate the role of the marginally force-free
    equatorial boundary condition.  We find this boundary condition plays an important role in
    shaping the BH magnetosphere in various aspects, including the shape of the light surface, the near-horizon field
    line configuration and the source of the Poynting flux.
    We also propose an algorithm for numerically solving the Grad-Shafranov equation and self-consistently imposing the marginally force-free equatorial
    condition. As a result, we find a good agreement between our numerical solutions and the high-accuracy FFE simulations.
    We also discuss the applicability of the marginally force-free boundary condition and the numerical algorithm
    proposed in this paper for general magnetic field configurations.
\end{abstract}

\maketitle

\section{Introduction}

The Blandford-Znajek (BZ) mechanism \cite{Blandford1977} is believed to be
one of most efficient ways to extract rotation energy from spinning black holes (BHs),
which operates in BH systems on all mass scales, from the stellar-mass BHs of gamma-ray
bursts to the supermassive BHs of active galactic nuclei. In the past decade,
we have studied the BZ mechanism from different approaches and the cross-check among
these different approaches has facilitated substantial progress in understanding the underlying
detailed physics. Taking the simple monopole magnetic field
configuration as an example, the solutions obtained from different approaches
are in quantitative agreement,  see e.g. \cite{Komissarov2001,Komissarov2004,Komissarov2004e, McKinney2004}
for general relativistic magnetohydrodynamic simulations,
\cite{Tanabe2008,Pan2015,Pan2015b, Gralla2014, Gralla2015, Penna2015, Grignani2018} for analytic solutions
and \cite{Contopoulos2013,Nathanail2014,Mahlmann2018} for numerical solutions.

But for other magnetic field configurations, there is no such good agreement, e.g.,
different approaches do not even reach a consensus on the solution uniqueness for the uniform field configuration.
Several force-free electrodynamics (FFE) simulations \cite{Komissarov2004e, Komissarov2005,
Komissarov2007,Palenzuela2010,Paschalidis2013,Yang2015,Carrasco2017} have been done and
the BH magnetospheres in these simulations all settle down to a steady state with similar final field configuration,
which is an indicator for solution uniqueness.

From the viewpoint of numerical solutions, the structure of a BH magnetosphere in axisymmetric and
steady state is governed by the Grad-Shafranov (GS) equation, which is a second-order differential equation of
the magnetic flux $\Ap(r,\theta)$, with two eigenfunctions $I(\Ap)$ and $\Omega(\Ap)$ to be
determined. For common field configurations, the two eigenfunctions are determined by requiring the
magnetic field line smoothly cross the light surfaces (LSs), where the GS equation degrades to be first-order.
But for the uniform field configuration, there exits only one LS, which is insufficient for determining
the two eigenfunctions. Following this argument, there should exist infinitely many solutions \cite{Nathanail2014}.
In addition, a family of analytic solutions were presented for slowly spinning BHs,
and no instability mode was found for any of these solutions \cite{Yang2015}. Therefore the solution stability
is not likely the explanation for the solution uniqueness.

To explain the discrepancy about the solution uniqueness from different approaches, Pan et al.  \cite{Pan2017}
proposed that the two eigenfunctions $\Omega(\Ap)$ and $I(\Ap)$ are connected by the radiation condition at infinity instead of
being independent, which was readily confirmed by recent high-accuracy FFE simulations done
by East and Yang \cite{East2018}. In addition, there are other interesting features in the structure of the BH
magnetosphere showing up in the simulations, e.g., an equatorial current sheet naturally develops
with the ergosphere, and the magnetic dominance marginally loses on the current sheet, i.e. $B^2-E^2\approx 0$.

Motivated by these simulation results, we revisit the uniform field solution and investigate the role of the
marginally force-free equatorial boundary condition in the BH magnetosphere structure.
We find the qualitative properties of the BH magnetosphere structure, including the shape of the LS,
the near-horizon field line configuration, and the source of the Poynting flux, are attributed to the marginally
force-free equatorial boundary condition without invoking the GS equation.
We also  propose an algorithm for numerically solving the GS equation and
self-consistently imposing the the marginally force-free equatorial boundary condition.
As a result, we find our numerical solutions are in good agreement with the FFE simulations.

The paper is organized as follows.  In Section \ref{sec:basic}, we outline the basic governing equations.
In Section \ref{sec:uni_sol}, we clarify the radiation condition, boundary conditions and the numerical algorithm
for the uniform field solution. In Section \ref{sec:discussion}, we generalize the discussion
to  more field configurations. Summary is given in Section \ref{sec:summary}.
Throughout this paper, we use the convention $G=c=M=1$ unless otherwise specified,
where $M$ the mass of the BH.

\section{basic equations}
\label{sec:basic}
In this paper, we adopt the Kerr-Schild
coordinate with the line element
\[
\begin{aligned}
ds^2 =
&-\left( 1-\frac{2r}{\Sigma} \right)dt^2 + \left( \frac{4
r}{\Sigma} \right) dr dt + \left(1+\frac{2r}{\Sigma} \right) dr^2 \\
&+ \Sigma d\theta^2 - \frac{4 a r \sin^2\theta}{\Sigma} d\phi dt
- 2 a \left(1+\frac{2r}{\Sigma}\right) \sin^2\theta d\phi dr     \\
& + \frac{\beta}{\Sigma} \sst d\phi^2
\end{aligned}
\]
where $\mu\equiv\ct$, $\Sigma = r^2 + a^2 \mu^2$, $\Delta = r^2 -2r + a^2$,
$\beta = \Delta\Sigma + 2r(r^2 + a^2)$, and $a$ is the dimensionless BH spin.
In the force-free approximation, electromagnetic energy greatly exceeds that of matter.
Consequently, the force-free magnetospheres is governed by energy
conservation equation of electromagnetic field, or
conventionally called as the GS equation.
In the Kerr spacetime,
the axisymmetric and steady GS equation can be written in a compact form \cite{Pan2017}
\ben
\label{eq:GSg}
&&\phantom{+}
 \left[\Arr + \frac{\sst}{\Delta}\Amm \right]  \mathcal K(r,\theta; \Omega )\nn \\
&&
+\left[\Ar \partial_r^\Omega  +  \frac{\sst}{\Delta}\Am \partial_\mu^\Omega\right] \mathcal K(r,\theta; \Omega ) \nn \\
&&
+ \frac{1}{2}\left[\Ar^2 + \frac{\sst}{\Delta}\Am^2\right]  \Omega' \partial_\Omega \mathcal K(r,\theta; \Omega )\nn \\
&&
- \frac{\Sigma}{\Delta}II' = 0 \ ,
\een
where the LS function
\be
\label{eq:ls}
\mathcal K(r,\theta; \Omega )= \frac{\beta}{\Sigma}\Omega^2 \sst
-\frac{4ra}{\Sigma}\Omega \sst
-\left(1-\frac{2r}{\Sigma}\right),
\ee
the primes designate  derivatives with respect to $A_\phi$.
$\partial_i^\Omega (i=r, \mu)$  denotes the partial derivative
with respect to coordinate $i$ with $\Omega$ fixed, and $\partial_\Omega$ is the derivative with
respect to $\Omega$. {\bf The GS equation degrades to first order on the LS,
where the LS function $\mathcal K(r,\theta; \Omega )$ vanishes.}

\section{Uniform field solution}
\label{sec:uni_sol}

\subsection{Solution uniqueness and radiation condition}
For common field configurations, there exists two LSs where
the LS function vanishes and the GS equation degrades from second order to first order. As proposed
by Contopoulos et al. \cite{Contopoulos2013}, one can adjust the two eigenfunctions $\Omega(\Ap)$
and $I(\Ap)$ enabling field lines smoothly cross the two LSs, then the solution
$\{\Omega(\Ap), I(\Ap), \Ap(r,\theta)\}$  is uniquely ensured. But for the vertical field lines, their
exists only one LS, which is insufficient for determining two eigenfunctions.
In this case, many solutions are expected \cite{Nathanail2014, Mahlmann2018}, but the many-solutions scenario is in
conflict with several previous FFE simulations \cite{Komissarov2004e, Komissarov2005,
Komissarov2007,Palenzuela2010,Paschalidis2013,Yang2015,Carrasco2017}.
To explain the discrepancy on the uniqueness of uniform field solution,
Pan et al. \cite{Pan2016a, Pan2017} proposed that the two eigenfunctions are not independent;
instead, they are related by the radiation condition at infinity, {\bf which is formulated as
$\hat E_\theta = \hat B_\phi$, with $\hat E_\theta$ and $\hat B_\phi$ being the $\theta$
component of electric field and $\phi$ component of the magnetic field measured by zero-angular-momentum-observers, respectively.
As for the uniform field solution, the radiation condition is explicitly expressed as }
\be I = 2\Omega A_\phi, \label{eq:rad}\ee
which has been readily confirmed by recent high-accuracy FFE simulations \cite{East2018}.
Combining with suitable boundary conditions, we expect a unique uniform field solution as indicated
by the previous FFE simulations.

\subsection{Boundary conditions}

The boundary conditions at infinity (inner infinity $r=r_+$ and outer infinity $r\rightarrow\infty$)
and on the polar axis can be simply set as
\be
\begin{aligned}
\Ar|_{r=r_+, \infty} &= 0, \\
A_\phi|_{\mu = 1} &= 0,\\
\end{aligned}
\ee
where $r_+$ is the radius of the event horizon,
while the equatorial boundary condition is more uncertain until recent high-accuracy simulations
come out showing that there exists an equator current sheet within the ergosphere where the
magnetic dominance marginally loses, i.e., $(B^2-E^2)/B_0^2$ goes to a small positive value
as approaching the current sheet, where $B_0$ is the uniform field strength
at infinity  \cite{East2018}.\footnote{The marginally force-free equatorial boundary condition
is not a unique feature of BH magnetospheres, which is also found in dissipative pulsar magnetospheres
\cite[see e.g.][]{Gruzinov2008, Gruzinov2011}.}
Motivated by the simulation results,
we choose the following equatorial boundary condition in our numerical solutions,
\begin{subequations}
\begin{align}
    \Am(\mu = 0, r > 2) &= 0, \label{eq:bc2}\\
    B^2-E^2 (\mu = 0, r_+ \leq r \leq 2) &=0.\label{eq:bc3}
\end{align}
\end{subequations}
In fact, Equation (\ref{eq:bc3})
is neither a Dirichlet nor a Neumann boundary condition, since
\be
\begin{aligned}
&B^2-E^2 \label{eq:B2mE2}\\
= & \frac{1}{\Sigma \sst} \left[ -\mathcal{K} \left(\Ar^2 +\frac{\sst}{\Delta}\Am^2 \right)+\frac{\Sigma}{\Delta}I^2\right],
\end{aligned}
\ee
which involves both derivatives $\Am$ and $\Ar$ on the boundary.
As we will see later, it is numerically non-trivial to impose this boundary condition in computation.

We note a coordinate singularity $1/\Delta$ in the expression of $B^2-E^2$.
To avoid possible numerical difficulty, we use the prescription
\be
\label{eq:nbc}
\int_{A_\phi^{\rm HE}}^{A_\phi^{\rm EE}} \left(\frac{B^2-E^2}{B^2+E^2}\right)^2 dA_\phi \Bigg/ \left(A_\phi^{\rm EE}- A_\phi^{\rm HE}\right)  < 10^{-3},
\ee
in our computation, as a proxy of the marginally force-free equatorial boundary condition (\ref{eq:bc3}),
where $\AHE$ and $\AEE$ are the magnetic flux enclosed by the horizon and by the ergosphere, respectively;
``HE" and ``EE" are short for Horizon-Equator and Ergosphere-Equator, respectively.
For definiteness, we choose $B^2+E^2$ to be the energy density measured by zero-angular-momentum-observers.
Explicitly, we have
\be
\begin{aligned}
 & B^2+E^2   \\
=& \frac{1}{\Sigma \sst} \left[ \left(\mathcal{K}+\frac{\Delta\Sigma}{\beta} \right)
 \left(\Ar^2 +\frac{\sst}{\Delta}\Am^2 \right)
 +\frac{\Sigma}{\Delta}I^2\right] .
\end{aligned}
\ee

\subsection{Generic properties of the BH magnetosphere structure}
\label{subsec:features}
Before delving into the details of numerically solving the GS equations,
here we point out that from the radiation condition (\ref{eq:rad}) and the marginally
force-free boundary condition (\ref{eq:bc3})
themselves contain rich information about the BH magnetosphere structure.

Let's first find out where the LS intersects with the equator, $r_{\rm LS}|_{\mu=0}$.
{\bf On this point $r_{\rm LS}|_{\mu=0}$ where the LS function $\mathcal K$ vanishes,
$I$ must also vanish for satisfying the marginally force-free boundary condition (see Equation [\ref{eq:B2mE2}]),
which in turns indicates a vanishing angular velocity $\Omega$ from
the radiation condition (\ref{eq:rad}), i.e., $\Omega(\mu=0, r= r_{\rm LS}|_{\mu=0}) =0$.
Plugging $\Omega(\mu=0, r= r_{\rm LS}|_{\mu=0}) =0$ back into $\mathcal K = 0$,
we obtain $r_{\rm LS}|_{\mu=0}=2$, i.e.
to satisfy the boundary condition (\ref{eq:bc3}), the LS must intersect the equator at $r=2$, which also
justifies our choice of equatorial boundary conditions (\ref{eq:bc2},\ref{eq:bc3}) . }

From above analysis, we expect several generic properties in the magnetospheres that are independent
of the GS equation:
(1) the LS runs from $r=r_+$ to $r=2$ as $\theta$ varies from $0$ to $\pi/2$;
(2) since $I$ vanishes at $r_{\rm LS}|_{\mu=0}$, we expect no current sheet within the magnetosphere except the equatorial current
sheet extending from $r_+$ to $2$, which gives rise to a cusp ($\Am \neq 0$) to the equatorial magnetic field lines;
(3) magnetic field lines entering the ergosphere end up either on the horizon or on the equatorial current sheet,
both of which carry electric current and therefore Poynting flux (see \cite{Punsly1990} for a physical realization of
equatorial current sheet sourcing Poynting flux).

With the guidance of the qualitative properties above, we now proceed to numerically
solve the GS equation and quantify these properties.

\subsection{Numerical method}
\label{subsec:algorithm}

In our computation, we define a new radial coordinate $R=r/(1+r)$, confine our
computation domain $R\times \mu$ in the region $[R(r_+), 1]\times [0,1]$,
and implement a uniform $512\times 64$ grid.
We aim to find a pair of $\Omega(A_\phi)$ and $I(A_\phi)$ satisfying the radiation condition (\ref{eq:rad})
and enabling field lines smoothly crossing the LS,
and suitable normal derivative $\Am(\mu =0, r_+\leq r \leq 2)$ on the equator
guaranteeing the boundary condition (\ref{eq:bc3}).

The numerical algorithm of searching for the desired eigenfunctions and the equatorial
boundary condition $\{\Omega(A_\phi), I(A_\phi), \Am(\mu =0, r_+\leq r \leq 2)\}$
is detailed in the following steps.

1. We choose an initial guess for the field configuration, eigenfunctions
$\{ \Omega(\Ap), I(\Ap)\}$
and equatorial boundary condition  as follows
\begin{subequations}
\begin{align}
    \Ap &= \frac{B_0}{2}r^2 \sst,  \label{eq:gss1} \\
    \Omega & = 0.5\WH\left(1-\Ap/\AHE\right),  \label{eq:gss2}\\
    I & = \WH \Ap\left(1-\Ap/\AHE\right) , \\
    \Am(\mu =0, r_+\leq r \leq 2) & = - (r/r_+)^3, \label{eq:gss3}
\end{align}
\end{subequations}
where $\WH = a/2r_+$ is the angular velocity of the BH.

2. We evolve the GS equation (\ref{eq:GSg}) using the well-known relaxation method \cite{Press1987}
 and adjust $II'(\Ap)$ until  field lines smoothly cross the LS
\cite[see e.g.][for more details]{Contopoulos2013, Nathanail2014, Pan2017, Mahlmann2018}.

3. Usually the current $I$ found in Step 2 neither satisfies the radiation condition (\ref{eq:rad})
nor guarantees the boundary condition (\ref{eq:bc3}). We  adjust
$\Am(\mu = 0, r_+ \leq r\leq 2)$ as follows,
\be
\label{eq:bc_crt}
 A_{\phi,\mu}|_{\rm new}  = A_{\phi,\mu}|_{\rm old} + \zeta_1\times[2\Omega\Ap(2\Omega\Ap)' -II'],
\ee
where $\zeta_1$ is an empirical step size. For each new $A_{\phi,\mu}$, we repeat Step 2 and iterative
correction (\ref{eq:bc_crt}) until $\Am(\mu = 0, r_+ \leq r\leq 2)$ converges, i.e. the condition $2\Omega\Ap(2\Omega\Ap)' = II'$ is
achieved for $A_\phi\in (A_\phi^{\rm HE}, A_\phi^{\rm EE})$.

4. The remaining task is to adjust  $\Omega(0< \Ap < \AHE)$ enabling  the radiation condition (\ref{eq:rad})
 for $A_\phi \in (0, A_\phi^{\rm HE})$ and  to adjust $\Omega(\AHE \leq \Ap \leq \AEE)$
 enabling the boundary condition (\ref{eq:bc3}) for $\Ap \in (\AHE, \AEE)$.
 The first part is straightforward, i.e.,
 \be
2\Ap\Omega_{\rm new}  = I|_{0 < \Ap < \AHE},
 \ee
 and the second part can be realized by iterative correction
 \be
2\Ap(\Omega_{\rm new} - \Omega_{\rm old}) =  - \zeta_2\times\Delta (B^2-E^2)|_{\mu=0, r_+\leq r \leq 2},
 \ee
 where $\zeta_2$ is again an empirical step size, and we have multiplied factor $\Delta$
 in the correction term to avoid numerical difficulty in the vicinity of the event horizon.
 To eliminate unphysical discontinuity in the angular
 velocity at $\AHE$, we fit $\Omega_{\rm new}(\Ap)$ on the whole range
 $(0, \AEE)$ via a fifth-order polynomial.

 5. For the new angular velocity $\Omega_{\rm new}(\Ap)$ obtained in Step 4, we repeat Step 2 to Step 4,
 until both the radiation condition (\ref{eq:rad}) and the numerical prescription (\ref{eq:nbc}) for the boundary condition (\ref{eq:bc3}) is satisfied.

\subsection{Numerical results}

\begin{figure}
\includegraphics[scale=0.6]{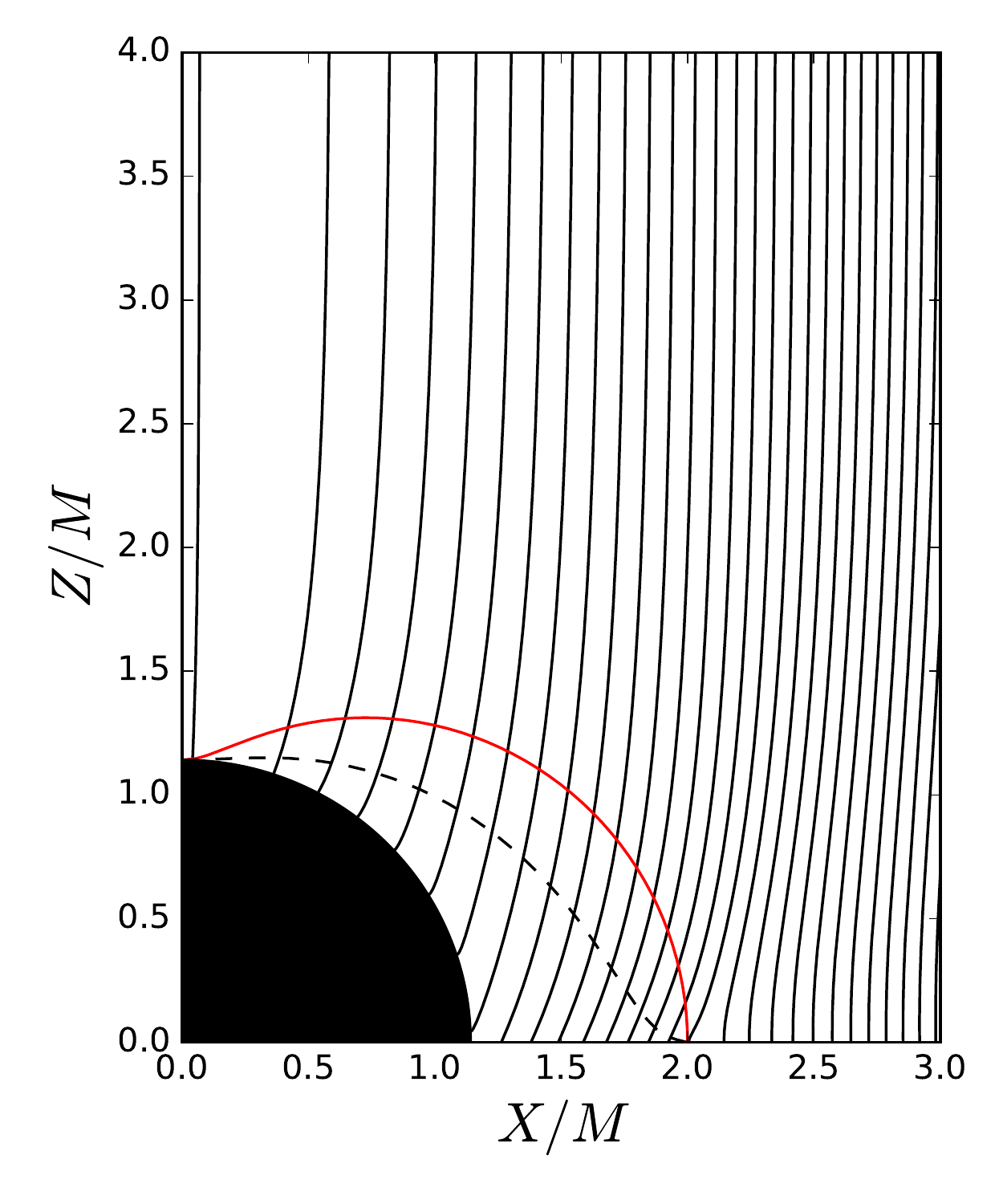}
\caption{\label{fig:field_lines} The configuration of field lines for the magnetosphere of a Kerr BH with spin $a=0.99$,
where the solid/red line is the ergosphere and the dashed/black line is the LS, both of which intersect with
the equator at $r=2 M$. }
\end{figure}

In Figure \ref{fig:field_lines}, we plot the magnetic field lines enclosing a BH with spin $a =0.99$ as an example,
which explicitly displays the properties we anticipated in Section \ref{subsec:features} and agrees with
the simulation results in detail \cite{East2018}.

\begin{figure}
\includegraphics[scale=0.7]{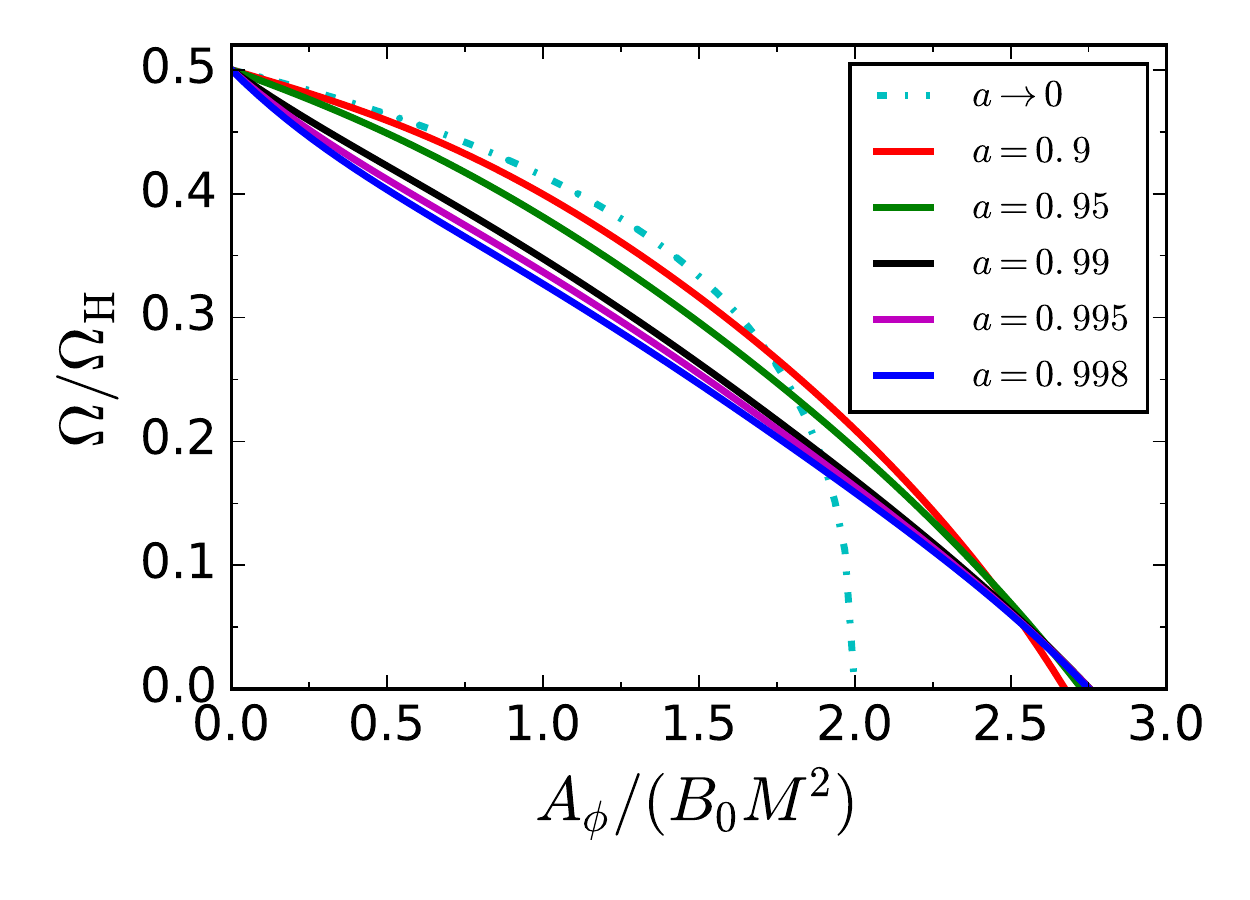}
\includegraphics[scale=0.7]{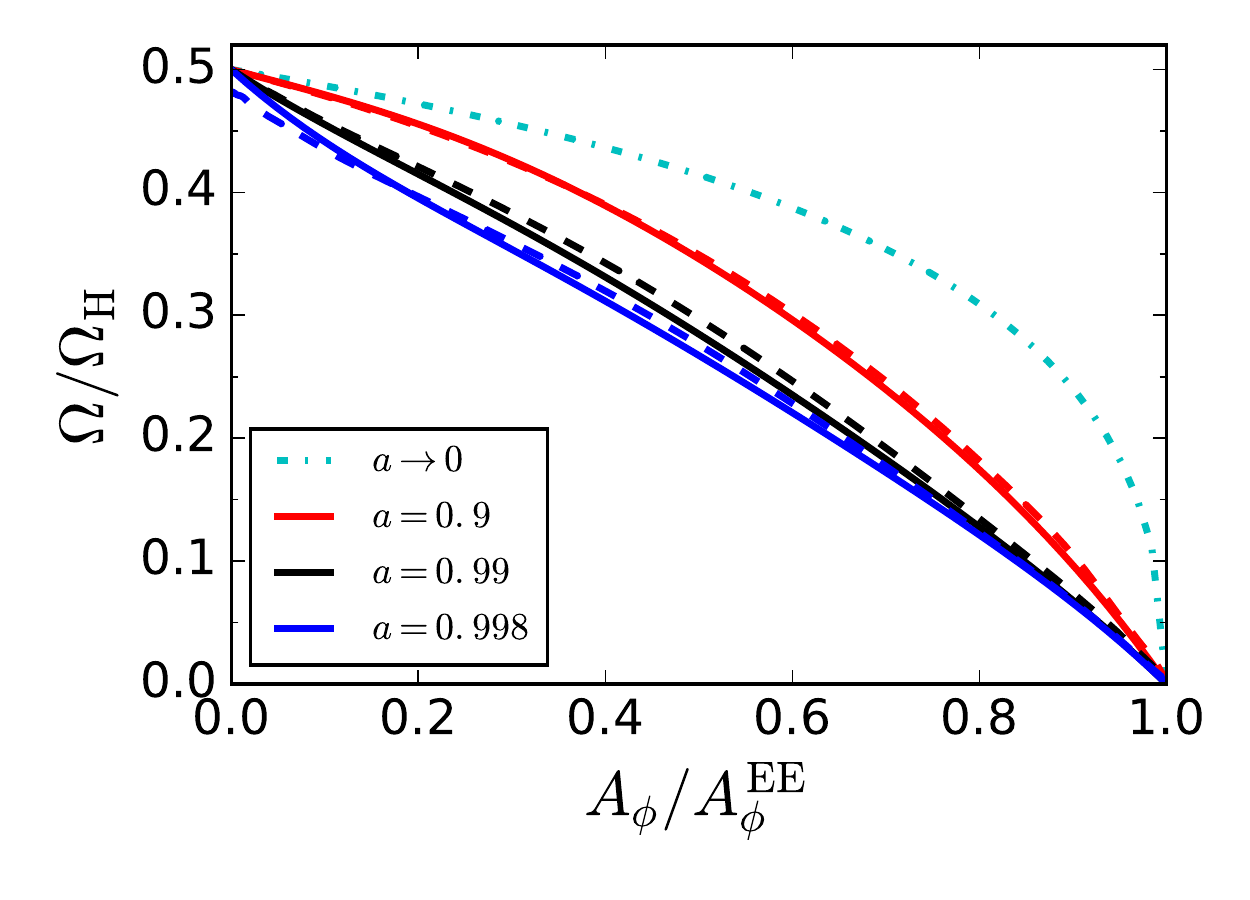}
\caption{\label{fig:omega} Upper panel: the angular velocity  $\Omega(\Ap)$ for different BH spins
obtained from our numerical solutions.
Lower Panel: comparison of our numerical results (solid lines) with the simulation results of
Ref. \cite{East2018} (dashed lines). For reference, we also plot the leading order analytic solution in
dash-dotted lines.}
\end{figure}

In Figure \ref{fig:omega}, we show the angular velocity function $\Omega(A_\phi)$ for different BH spins and
compare it with the counterpart obtained from the simulations \cite{East2018}.
For reference, we also plot the leading-order analytic solution in the slow-rotation
limit \cite{Beskin2013, Pan2014, Gralla2015, East2018},
\be
    \Omega = \WH\frac{\sqrt{1-\psi}} {1+\sqrt{1-\psi}},
\ee
where $\psi = \Ap/(2B_0M^2)$.
From our numerical solutions, we find the magnetic flux entering the ergosphere $\AEE$ increases with
the BH spin and approaches $2.75 B_0 M^2$ for extremal spins (upper panel of Figure \ref{fig:omega}),
which is about $\approx 5\%$ lower than the simulation result (Figure 3 in Ref. \cite{East2018}),
while the angular velocity $\Omega$ as a function
of normalized magnetic flux $\Ap/\AEE$ is in agreement with the
simulation results to high precision.\footnote{We have done a test and
find that the $\approx 5\%$ difference in $\AEE$ is not arising from the slightly
different equatorial boundary conditions used in this work and found from East and Yang's simulations.
The difference is more likely due to the relative numerical bias between the two algorithms. }

With the angular velocity $\Omega(\Ap)$ obtained, the energy extraction rate from the BH is given by
\be
\label{eq:Edot}
\dot E = 4\pi \int_0^{\AEE} \Omega \times I \ d \Ap.
\ee
It is straightforward to obtain the energy extraction rate in the slow-rotation limit
\be
\label{eq:lowEdot}
\dot E = 128\pi \left(\frac{17}{24}-\ln 2\right) B_0^2 M^4 \WH^2.
\ee

\begin{figure}
\includegraphics[scale=0.7]{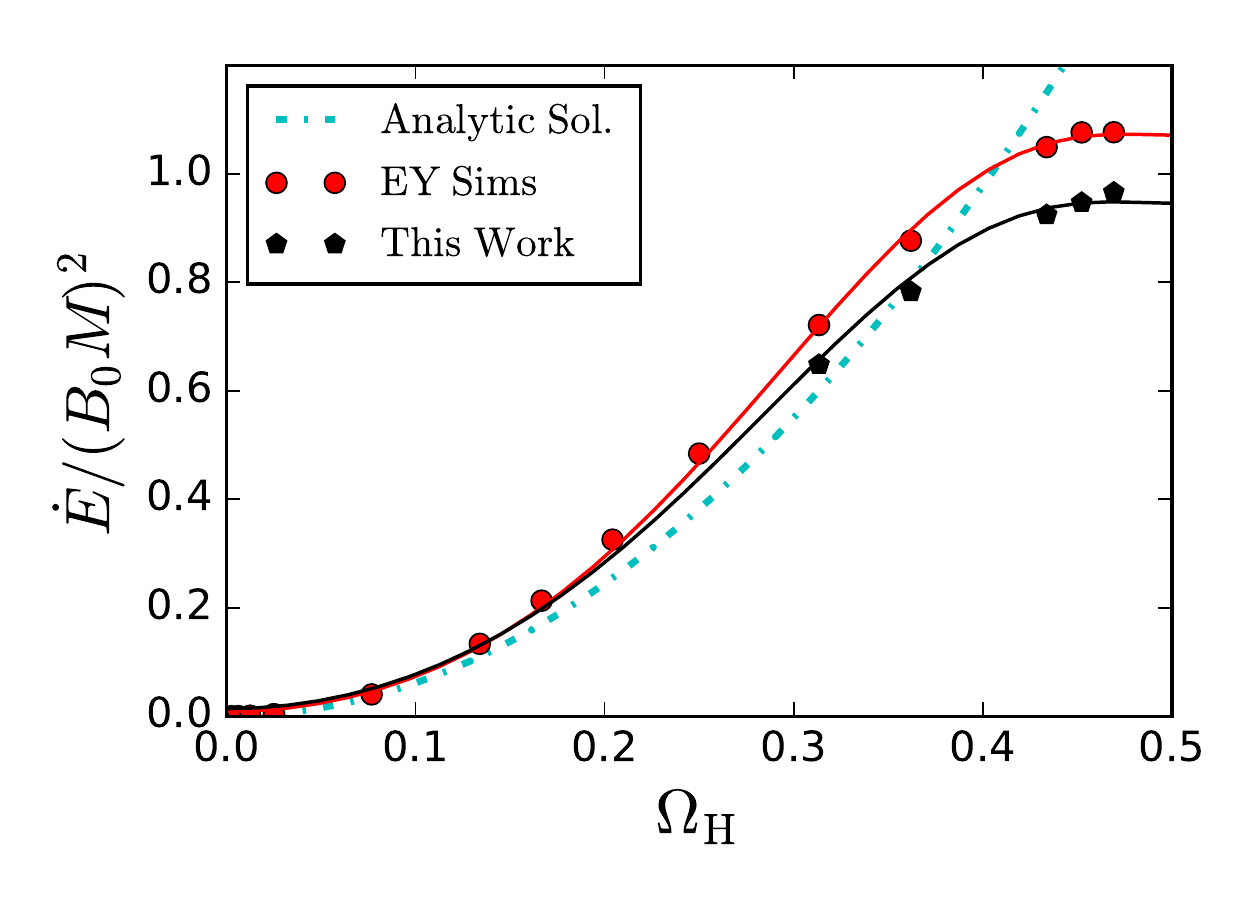}
\caption{\label{fig:Edot} Comparison of the energy extraction rates $\dot E(\WH)$ obtained from
three different approaches: the leading-order analytic solution (\ref{eq:lowEdot}),
our numerical solutions and  the high-resolution force-free simulations \cite{East2018}.}
\end{figure}

In Figure \ref{fig:Edot}, we compare the energy extraction rates $\dot E(\WH)$
derived from our numerical solutions with East and Yang's simulation results
\cite{East2018}, where the data points are taken from either the simulations
or our numerical solutions, while the solid lines are corresponding polynomial
fitting curves which we require to approach Equation (\ref{eq:lowEdot}) for small
spins and to be flat for extremal spins. As expected, our energy extraction
rate $\dot E(\WH)$ is $\approx 10\%$ lower than the corresponding simulation results,
due to the $\approx 5\%$ smaller magnetic flux $\AEE$.

To summarize, the uniform field solution is indeed unique as double confirmed by the high-accuracy FFE
simulations and by our numerical solutions. The  structure of the BH magnetosphere
is largely shaped by the radiation condition and the marginally force-free
equatorial boundary condition.

\section{Discussion}
\label{sec:discussion}

\subsection{Application to general field configurations}
In real astrophysical environment, we expect the field lines far away from the central BH
are more close to parabolas instead of being strictly vertical. In several previous studies of such field
configurations \cite[e.g.][]{Tchekhovskoy2010,Nathanail2014,Mahlmann2018},
due to lacking knowledge of the equatorial boundary condition, the
equator within the ergosphere was intentionally excluded out of the computation domain by manually
introducing a ``wall" extending from the horizon-equator intersection to infinity. Such
simplification obviously misses magnetic field lines rooting on the
equatorial current sheet, which contribute about half of the total Poynting flux for extremal spins in
the case of uniform field configuration.

Due to the resemblance of near-horizon field lines in the two cases,
it is reasonable to expect an equatorial current develops within the ergosphere,
where the magnetic dominance loses, therefore the marginally force-free boundary
condition (\ref{eq:bc3}) should also be a good work approximation for studying
the BH magnetosphere embedded in parabolic magnetic field lines.
It is straightforward to solve the GS equation and self-consistently
impose the marginally force-free boundary condition following the
algorithm detailed in Section \ref{subsec:algorithm}.

Though we do not numerically solve the GS equation for the general parabolic field configurations,
the qualitative properties we summarized in Section \ref{subsec:features} also apply here,
since these properties are the consequence of the radiation condition and the marginally force-free equatorial boundary condition,
while the GS equation only serves to quantify them.

\subsection{Near-horizon magnetic field lines}
In a previous study \cite{Pan2016a}, we made a claim that ``in the steady axisymmetric force-free magnetosphere
around a Kerr BH, all magnetic field lines that cross the infinite-redshift surface must intersect the event horizon".
This claim is based on the radiation condition
\be
\label{eq:Grad}
I = \Omega \times \mathcal F(\Ap),
\ee
and the assumption of no current sheet within the ergosphere,
where the function $\mathcal F(\Ap)$ is of $\mathcal O(\Ap)$ and  is field configuration dependent.
The basic logic for obtaining the claim above is as follows.
The angular velocity $\Omega$ must be nonzero for all field lines entering the ergosphere due to the frame-dragging effect;
as a result, $I$ must be nonzero for these field lines according to the radiation condition.
If there is a field line entering the ergosphere and crossing the equator, the electric current either flows
towards the equator from both the $+z$ and $-z$ side, or flows away from the equator to infinity in
both the $+z$ and $-z$ direction. For each case, the charge conservation is violated
if there exists no equatorial current sheet.

However, the high-accuracy FFE simulations show that an equatorial current sheet inevitably develops within the ergosphere, where
the force-free condition marginally breaks down. Therefore the above claim should be generalized as
``in the steady axisymmetric force-free magnetosphere
around a Kerr BH, all magnetic field lines that cross the infinite-redshift surface
must intersect the event horizon or end up on the equatorial current sheet". Specifically,
this claim excludes the existence of field lines entering the ergosphere and crossing the
equator vertically.

\section{Summary}
\label{sec:summary}

In the force-free limit, the structure of steady and axisymmetric BH magnetosphere is governed by
the GS equation, which is a second-order differential equation about the magnetic flux $\Ap$, with two
eigenfunctions $\Omega(\Ap)$ and $I(\Ap)$ to be determined. For common field configurations, there exists
two LSs on which the GS equation degrades to be first-order, and the two eigenfunctions are determined
by the requirement that magnetic field lines should smoothly cross the two LSs.
For the uniform field configuration, there is only one
LS, which is insufficient for determine both  $\Omega(\Ap)$ and $I(\Ap)$.
Therefore the solution uniqueness of the uniform field configuration has been a controversial problem.
To tackle this problem, we  proposed that the two functions are related by the radiation condition (\ref{eq:rad}),
instead of being independent \cite{Pan2017}, which was readily confirmed by recent high-accuracy
FFE simulations \cite{East2018}. In addition, these simulations also provide a close look at the
equatorial boundary condition: an equatorial current sheet develops within the ergosphere and the magnetic dominance marginally loses, i.e. $B^2-E^2\approx 0$.

Motivated by these simulation results, we revisit the problem of the uniform field solution in this paper.
We find the radiation condition (\ref{eq:rad}) and the marginally force-free boundary
condition (\ref{eq:bc3}) are rather informative, which dictate the BH magnetosphere structure
in various aspects, including the shape of the LS, the near-horizon field line configuration and
the source of BZ flux (see Section \ref{subsec:features} for details). {\bf Especially we find the LS
intersects with the ergosphere at the equator, which was also observed in previous simulations \cite[e.g.][]{Carrasco2017, East2018}
and now we understand its underlying physics: the radiation condition and the marginally force-free condition.}
Other than these qualitative properties,
we also propose an algorithm for numerically solving the GS equation and consistently imposing
the marginally force-free equatorial boundary condition. As a result, we find a good agreement
between our numerical solutions with the high-accuracy FFE simulations.

In realistic astrophysical environment, we expect the magnetic field lines far away from the central BH
are more close to be parabolic instead of being strictly vertical. However, we also expect the marginally
force-free equatorial boundary condition to be a good working approximation for studying the parabolic field configurations,
due to the resemblance of the near-horizon field configurations in the two cases.
Though we do not numerically solve the GS equation for the parabolic configurations in this paper, the qualitative
properties of the uniform field solution summarized in Section \ref{subsec:features} also apply here,
since these properties are dictated by the radiation condition
and the marginally force-free boundary condition, while the GS equation only serves to quantify them.

\acknowledgements
ZP thanks William East and Huan Yang for stimulating discussions and sharing their simulation results.
ZP also thanks Cong Yu and Lei Huang for reading through a previous version of this manuscript and providing useful suggestions.
ZP was supported by the UC Davis Dissertation Year Fellowship when this work was started.
This research was also supported by Perimeter Institute for Theoretical Physics.
Research at Perimeter Institute is supported by the Government of Canada
through the Department of Innovation, Science and Economic Development Canada
and by the Province of Ontario through the Ministry of Research, Innovation and Science.

\bibliography{ms}

\end{document}